\documentstyle[12pt,epsfig]{article}
\begin{document}
\newcommand{\beq}{\begin{equation}}
\newcommand{\eeq}{\end{equation}}
\newcommand{\bea}{\begin{eqnarray}}
\newcommand{\eea}{\end{eqnarray}}
\newcommand{\lam}{\lambda}
\newcommand{\GE}{\gamma_{e}}
\newcommand\MSb{$\overline{\mbox{MS}}$}
\newcommand{\gsim}{\raisebox{-0.07cm}{$\:\stackrel{>}{{\scriptstyle
 \sim}}\: $} }
\newcommand{\lsim}{\raisebox{-0.07cm}{$\:\stackrel{<}{{\scriptstyle
 \sim}}\: $} }
\setlength{\baselineskip}{0.58cm}
\setlength{\parskip}{0.35cm}
\begin{titlepage}

\noindent
{\tt hep-ph/0010146} \hfill INLO-PUB 10/00 \\
\hspace*{\fill} October 2000 \\
\vspace{1.5cm}
\begin{center}
\large
{\bf Next-to-Next-to-Leading Logarithmic Threshold Resummation} \\
\vspace{0.1cm}
{\bf for Deep-Inelastic Scattering and the Drell-Yan Process} \\
\large
\large
\vspace{2.8cm}
A. Vogt \\
\vspace{1.2cm}
\normalsize
{\it Instituut-Lorentz, University of Leiden \\
\vspace{0.1cm}
 P.O. Box 9506, 2300 RA Leiden, The Netherlands} \\
\vfill
\large
{\bf Abstract} \\
\end{center}
\vspace{-0.3cm}
\normalsize
The soft-gluon resummation exponents $G^N$ in moment space are 
investigated for the quark coefficient functions in deep-inelastic 
structure functions and the quark-antiquark contribution to the 
Drell-Yan cross section $d\sigma/dM$. Employing results from two-
and three-loop calculations we obtain the next-to-next-to-leading
logarithmic terms $\alpha_s (\alpha_s \ln N)^n$ of $G^N$ to all
orders in the strong coupling constant $\alpha_s$. These new
contributions facilitate a reliable assessment of the numerical effect 
and the stability of the large-$N$ expansion. 
\vspace*{0.5cm}
\normalsize

\end{titlepage}

\noindent
Deep-inelastic lepton-hadron scattering (DIS) and Drell-Yan (DY)
lepton pair production in hadronic collisions are among the processes
best suited for probing the short-distance structure of hadrons.
The subprocess cross sections (coefficient functions) for these 
processes in perturbative QCD receive large logarithmic corrections, 
originating from soft-gluon radiation, at large values of the scaling 
variables $x$ corresponding to large Mellin moments $N$. These 
corrections have been resummed \cite{sglue1,sglue2} to all orders in 
the strong coupling constant $\alpha_s$ up to next-to-leading 
logarithmic accuracy (technically defined in \mbox{eq.~(\ref{GNexp})} 
below). 
As the next-to-leading contributions are often hardly suppressed 
against the leading terms, it is important to extend the resummation to 
the next-to-next-to-leading logarithms. In this paper we present 
corresponding results for the DIS structure functions $F_{1,2,3}(x,Q^2)$
(where $Q^2$ represents the resolution scale) and for the Drell-Yan 
cross section $d\sigma/dQ^2$ (where $Q$ stands for the invariant mass 
$M$ of the lepton pair). 

The soft-gluon resummation of the DIS ($P \equiv 1$) and DY ($P \equiv 
2$) quark coefficient functions $C_P^N (Q^2)$ in $N$-space is given by
\cite{sglue1,sglue2}
\beq
\label{cNres}
  C_P^N (Q^2) \: =\: g_{P,0}^{}(Q^2) \cdot \exp\, [G_P^N(Q^2)] 
  \: + \: {\cal O}(N^{-1}\ln^n N) \:\: .
\eeq
The functions $C^N\!$, $g_0^{}$ and $G^N$ also depend on the 
factorization scale $\mu_f$ and the renormalization scale $\mu_r$, a
dependence which we will often suppress for brevity. For the quantities 
under consideration the contributions $g_0$ collecting the 
$N$-independent terms are known up to second order in $\alpha_S$ from 
the two-loop calculations performed in refs.~\cite{softDY}. 
In the standard \MSb\ renormalization and factorization scheme employed
throughout this paper the exponents $G_P^N$ in eq.~(\ref{cNres}) can 
be written as 
\bea
\label{GNdec}
  G_1^N(Q^2) & \: = \: & \ln \Delta_q (Q^2,\mu_f^2) + \ln J_q(Q^2) 
    + \ln \Delta^{\rm int}_1(Q^2) \:\: , \nonumber \\[1mm]
  G_2^N(Q^2) & \: = \: & 2\, \ln \Delta_q (Q^2,\mu_f^2) + 
    \ln \Delta^{\rm int}_2(Q^2) \:\: .
\eea
Closely following the notations of ref.~\cite{CMN98}, the components 
entering eq.~(\ref{GNdec}) are
\beq
\label{dint}
  \ln \Delta_q (Q^2, \mu_f^2) \: = \: \int_0^1 \! dz \,
  \frac{z^{N-1}-1}{1-z} \,\int_{\mu_f^2}^{(1-z)^2 Q^2} 
  \frac{dq^2}{q^2}\, A(a_s(q^2)) 
\eeq
collecting of effects of soft-gluon radiation collinear to 
initial-state partons, 
\beq
\label{Jint}
  \ln J_q (Q^2) \: = \:  \int_0^1 \! dz \,\frac{z^{N-1}-1}{1-z}\,
  \left[ \int_{(1-z)^2 Q^2}^{(1-z) Q^2} \frac{dq^2}{q^2}\,
  A(a_s(q^2)) + B (a_s([1-z] Q^2)) \right] 
\eeq
taking into account collinear final-state radiation, and the 
process-dependent piece
\beq
\label{Dint}
  \ln \Delta^{\rm int}_P (Q^2) \: = \: \int_0^1 \! dz 
  \,\frac{z^{N-1}-1}{1-z} \, D_P(a_s([1-z]^2 Q^2)) 
\eeq
attributed to large-angle soft-gluon emissions. 
The integrands in eqs.~(\ref{dint})--(\ref{Dint}) are given by
\beq
\label{abexp}
  F(a_s) \: = \: \sum_{l=1}^{\infty} F_l\, a_s^l \: , \:\:\:
  F = A, \, B, \, D_P \:\: ,
\eeq
where we normalize the expansion parameter as $a_s = \alpha_s 
/(4\pi)\,$.
 
The constants $A_l$ in eq.~(\ref{abexp}) are the coefficients of 
the $1/[1-x]_+$ terms of $l$-loop quark-quark splitting functions 
$P_{qq}^{(l-1)}(x)$ --- this actually completes the \MSb\ definition of 
the coefficients $B_l$ and $D_{P,l}$ \cite{sglue2} (see also ref.\
\cite{BB}). Thus $A_1$ and $A_2$ are well-known, reading
\beq
\label{a12}
  A_1 \: = \: 4 C_F \: , \:\:\: 
  A_2 \: = \: 8 C_F \left[ \left( \frac{67}{18} - \zeta_2^{} \right) 
  C_A^{} - \frac{5}{9}\, N_f^{} \right] \:\: .
\eeq
Here $N_f$ denotes the number of effectively massless quark flavours, 
and the colour factors are $C_F = 4/3$ and $C_A = 3$ in QCD.
The exact expression for $P_{qq}^{(2)}(x)$ has not been completed yet. 
However, recently rather accurate approximations have been derived
\cite{NVplb} from the available partial results, most notably the 
lowest integer-$N$ moments \cite{moms1,moms2}. Together with the exact 
$N_f^2$ term determined in ref.~\cite{Gra1} the results of 
ref.~\cite{NVplb} yield\footnote
{The error estimate (\ref{a3}) has been derived in ref.~\cite{NVplb}
 under an assumption (no $\ln^3 (1-x)$ terms) on the form of 
 $P_{qq}^{(2)}(x)$. Abandoning this constraint does not lead to a 
 significant modification of eq.~(\ref{a3}).}
\beq
\label{a3}
  A_3 \: = \: (1178.8 \pm 11.5) - (183.95 \pm 0.85) N_f 
              - \frac{16}{27}\, C_f N_f^2 \:\: .
\eeq 
For $N_f = 3\,\ldots\, 5$ the (independent) errors in eq.~(\ref{a3}) 
can be combined to an overall uncertainty of $\pm 12$. For $\alpha_s < 
0.3$ this uncertainty amounts to less than 0.1\% of the total three-loop
value of $A(a_s)$. Finally the constants $B_1$ and $D_{P,1}$ in 
eq.~(\ref {abexp}) are given by \cite{sglue2}
\beq
\label{B1D1}
  B_1 \: = \: - P^{(0)}_{q,\delta} \: = \: - 3\, C_F \: , \:\:\: 
  D_{P,1} \: = \: 0 \:\: ,
\eeq
where $P^{(0)}_{q,\delta}$ is the coefficient of $\delta(1-x)$ in the
one-loop quark-quark splitting function. 
The second-order terms $B_2$ and $D_{P,2}$ will be discussed below. 
 
After the integrations in eqs.~(\ref{dint})--(\ref{Dint}) are performed,
the functions $G_P^N(Q^2)$ in eq.~(\ref{GNdec}) take the form
\beq
\label{GNexp}
  G^N(Q^2) \: = \: L\, g_1^{}(\lam) + g_2^{}(\lam) + a_s\, g_3^{}(\lam)
                   + \ldots 
\eeq
with $\, L = \ln N$, $\lam = \beta_0 a_s L\, $ and
\beq
\label{giexp}
  g_{i}^{}(\lam) \: = \: \sum_{k=1}^{\infty} g_{ik}^{}\, (a_s L)^k
  \:\: .
\eeq
The first term in eq.~(\ref{GNexp}), which depend on $A_1$ only
(see eq.~(\ref{g1n}) below), 
collects the leading logarithmic (LL) large-$N$ contributions 
$L (a_s L)^n$. The coefficients $A_2$ and $B_1$ determine the functions 
$g_2^{}$ resumming the  next-to-leading logarithmic (NLL) terms 
$(a_s L)^n$. These functions have been determined in refs.~\cite
{sglue1,sglue2}.  The next-to-next-to-leading logarithmic (NNLL) 
approximation includes $g_3^{}$ which is correspondingly fixed by 
$A_3$, $B_2$ and $D_{P,2}$.

In order to calculate $g_3$ it is convenient to introduce
\beq
\label{par1}
  X = 1 + a_s(\mu_r^2) \beta_0 \ln \frac{q^2}{\mu_r^2} 
\eeq
and to write the next-to-next-to-leading order coupling constant as
\bea
\label{arun}
  a_s(q^2) &\! =\! &
  a_s(\mu_r^2) \,\frac{1}{X} \: - \: 
  a_s^2(\mu_r^2) \,\frac{\beta_1}{\beta_0}\, \frac{\ln X}{X^2} 
  \nonumber \\[0.5mm] & & \mbox{} + \:
    a_s^3(\mu_r^2) \left( \frac{\beta_1^2}{\beta_0^2}\, 
    \frac{\ln^2 X - \ln X - 1 + X}{X^3} 
    + \frac{\beta_2}{\beta_0}\, \frac{1-X}{X^3} \right)
  \nonumber \\[1mm] & & \mbox{} + \:
  {\cal O} \Big( a_s^4(\mu_r^2) \Big[ a_s(\mu_r^2) 
    \ln(q^2/\mu_r^2) \Big]^n \Big) \:\: .
\eea
Here $\beta_0$, $\beta_1$ and $\beta_2$ are the coefficients of the
$\beta$-function of QCD up to three loops \cite{beta2}. After inserting 
eq.~(\ref{arun}) into eqs.~(\ref{dint}) and (\ref{Jint}), the inner 
integrations over $A(a_s)$ can be readily carried out. A straightforward
method to perform the $z$-integral is to write the integrand as an
infinite series in $\ln (1-x)$, to use 
\bea
  \int_0^1 \! dz\: \frac{z^{N-1}-1}{1-z}\, \ln^k (1-x) &\! = \! &
  \frac{(-1)^{k+1}}{k+1} \left\{ S_1^{k+1}(N) + \frac{1}{2}\, k(k+1)\,
   S_1^{k-1}(N)\, S_2(N) \right\}
  \nonumber \\[1mm] & & \mbox{} + {\cal O}(S_1^{k-2}) 
\eea
(cf.\ ref.~\cite{BK98}) with $S_a(N) = \sum_{j=1}^{N} 1/j^a$ and
\bea
  S_1(N) \: = \: \ln N + \gamma_e + {\cal O}(1/N) \: , \:\:\: 
  S_2(N) \: = \: \zeta_2 + {\cal O}(1/N) \:\: ,
\eea
and to re-assemble the expansions in the end. In this way we arrive at
\bea
\label{g1n}
  g_1^{\rm DIS}(\lam) &\! =\! & \frac{A_1}{\beta_0 \lam}\,
    \Big[ \lam + (1-\lam) \ln(1-\lam) \Big] \\[2mm]
\label{g2n}
  g_2^{\rm DIS}(\lam) &\! =\! & \mbox{}
    - \frac{A_1\GE - B_1}{\beta_0}\, \ln(1-\lam)
    + \frac{A_1 \beta_1}{\beta_0^3}\,
    \Big[ \lam + \ln (1-\lam) + \frac{1}{2}\ln^2(1-\lam) \Big]  
  \nonumber \\[1mm] 
  & & \mbox{} 
    - \frac{A_2}{\beta_0^2}\, \Big[ \lam + \ln(1-\lam)\Big]
    + \ln \left( \frac{Q^2}{\mu_r^2} \right) \frac{A_1}{\beta_0} 
    \ln (1-\lam) + \ln \left( \frac{\mu_f^2}{\mu_r^2} \right)
   \frac{A_1}{\beta_0} \,\lam 
\eea
and our new result 
\bea
\label{g3n}
  g_3^{\rm DIS}(\lam) &\! =\!\! & \mbox{} + 
  A_1 \left\{ \,\frac{1}{2} (\GE^2 + \zeta_2) \,\frac{\lam}{1-\lam}
      \, +\, \frac{\beta_1^2}{\beta_0^4}\, \frac{1}{1-\lam}
        \left[ \,\frac{1}{2} \ln^2 (1-\lam) + \lam \ln (1-\lam) +
        \frac{1}{2} \lam^2 \right] \right.
  \nonumber \\[0.5mm] & & \left. \mbox{} \quad\quad\quad
      -\, \frac{\beta_1 \GE}{\beta_0^2} \,\frac{1}{1-\lam}
        \Big[ \lam + \ln (1-\lam) \Big] 
      \, +\, \frac{\beta_2}{\beta_0^3} \,\bigg[ \,\frac{1}{2} 
        \frac{\lam^2}{1-\lam} + \ln (1-\lam) + \lam \bigg] \right\} 
  \nonumber \\[1mm] & & \mbox{} +
  A_2 \left\{ \frac{\GE}{\beta_0}\, \frac{\lam}{1-\lam}
      \, -\, \frac{\beta_1}{\beta_0^3}\, \frac{1}{1-\lam} \bigg[ 
        \ln (1-\lam) + \lam + \frac{1}{2} \lam^2 \bigg] \right\} \: +\: 
  \frac{A_3}{2\beta_0^2}\, \frac{\lam^2}{1-\lam}
  \nonumber \\[1mm] & & \mbox{} -
  B_{1} \left\{ \GE\, \frac{\lam}{1-\lam} 
      - \frac{\beta_1}{\beta_0^2}\, \frac{1}{1-\lam} 
        \Big[ \lam + \ln (1-\lam) \Big] \right\} \: - \:
  \frac{B_{2}}{\beta_0}\, \frac{\lam}{1-\lam} \: - \:
  \frac{D_{1,2}}{\beta_0}\, \frac{\lam}{1-2\lam}
  \nonumber \\[1mm] & & \mbox{} + 
  \ln \left( \frac{Q^2}{\mu_r^2} \right) \left\{ 
      \frac{A_1 \beta_1}{\beta_0^2}\, \frac{1}{1-\lam} \Big[ \lam + 
      \ln (1-\lam) \Big] + \bigg( B_{1} - A_1\GE -\frac{A_2}{\beta_0} 
      \bigg) \,\frac{\lam}{1-\lam} \right\}
  \nonumber \\[1mm] & & \mbox{} +
  \ln^2 \!\left( \frac{Q^2}{\mu_r^2} \right) \frac{A_1}{2}\,
      \frac{\lam}{1-\lam} \: + \: 
  \ln \left( \frac{\mu_f^2}{\mu_r^2}\right) \frac{A_2}{\beta_0}\, 
      \lam \: - \:
  \ln^2 \!\left( \frac{\mu_f^2}{\mu_r^2} \right) \frac{A_1}{2}\, \lam 
  \:\: .
\eea
The functions $g_{2,3}^{\rm DY}$ are obtained from eqs.~(\ref{g2n}) 
and (\ref{g3n}) by substituting $\lam \rightarrow 2\lam$, $\gamma_e 
\rightarrow 2\gamma_e$ and $\zeta_2 \rightarrow 4 \zeta_2$ in the terms 
with $A_i$, removing the terms with $B_l$ (recall eq.~(\ref{GNdec})), 
and replacing $D_{1,1}$ by $D_{2,1}$. The result corresponding to 
eq.~(\ref{g1n}) reads $g_1^{\rm DY}(\lam) = 2\, g_1^{\rm DIS}(2\lam)$.
The generalization of eqs.~(\ref{g1n})--(\ref{g3n}) to other processes 
involving eqs.~(\ref{dint})--(\ref{Dint}) is obvious.

Now we are ready to address the second-order coefficients $B_2$ and 
$D_{P,2}$. The first-order expansion coefficients $g_{31}^{\rm DIS}$ 
and $g_{31}^{\rm DY}$, as defined in eq.~(\ref{giexp}), are given by
\beq
  g_{31}^{(P)} \: = \: 1/2\: P^3 A_1 (\GE^2 + \zeta_2) \beta_0 
  + P^2 A_2 \GE - \delta_{P1} (B_2 - B_1 \GE \beta_0) - D_{P,2} 
\eeq
with $\delta_{kj} = 1$ for $k =  j$ and $\delta_{kj} = 0$ else.
On the other hand $g_{31}^{(P)}$ can be determined by expanding 
eqs.~(\ref{cNres}) and (\ref{GNexp}) to order $\alpha_s^2$ and 
comparing to the two-loop results of refs.~\cite
{softDY,MvNZP}, as done for the DIS case in ref.~\cite{av99}%
\footnote{Note that for $g_3$ the convention (\ref{giexp}) differs
 from that in ref.~\cite{av99}: there the second index $i$ in $g_{3i}$ 
 refers to the overall power of $\alpha_s$ in the expansion 
 (\ref{GNexp}) of $G^N$. Hence $g_{31}$ is called $g_{32}$ in 
 ref.~\cite{av99} etc.}.
The result for the Drell-Yan coefficient function reads
\bea
\label{g31DY}
  g_{31}^{\rm DY} &\! =\! & \mbox{} +
    C_F^{} C_A^{} \left( \frac{1616}{27} - 56\, \zeta_3^{} 
    - 32\,\zeta_2^{} \GE + \frac{176}{3}\, \GE^2
    + \frac{1072}{9}\, \GE \right) 
  \nonumber \\[1mm] & & \mbox{} -
     C_F^{} N_f^{} \left( \frac{224}{27} + \frac{32}{3}\,
    \GE^2 + \frac{160}{9}\, \GE \right) \:\: ,
\eea
yielding 
\beq
\label{D2DY}
  D_2^{\rm DY} = 
  C_F C_A \left( - \frac{1616}{27} + 56\,\zeta_3 + 
                 \frac{176}{3}\,\zeta_2 \right) \: + \:
  C_F N_f \left( \frac{224}{27} - \frac{32}{3}\,\zeta_2 \right) \:\: .
\eeq
This term has already been derived in ref.~\cite{CLS}, albeit without
explicitly attributing it to the $\alpha_s((1-z)^2 Q^2)$ contribution.
Note that, as it has to be for a $\Delta^{\rm int}$ contribution~\cite
{CMN98}, eq.~(\ref{D2DY}) does not contain an Abelian ($C_F^2$) term. 
We consider this as a first concrete check of the correctness of the 
setup (\ref{dint})-(\ref{Dint}) at the NNLL level.
Using eq.~(12) of ref.~\cite{av99}, the corresponding constraint for
the DIS case reads
\bea
\label{BD2DIS}
  B_2+ D_2^{\rm DIS} &\! =\! &
  C_F^2 \bigg( - \frac{3}{2} - 24\,\zeta_3 + 12\,\zeta_2 \bigg) \: + \: 
  C_F C_A \bigg( - \frac{3155}{54} + 40\,\zeta_3 
                 + \frac{44}{3}\,\zeta_2 \bigg) 
  \nonumber \\ & & \mbox{} + \:  
  C_F N_f \bigg( \frac{247}{27} - \frac{8}{3}\,\zeta_2 \bigg) 
  \: = \: 
  - P_{q,\delta}^{(1)} + \frac{1}{2}\, D_2^{\rm DY} - 7\, \beta_0 C_F
  \:\: ,
\eea
where $P_{q,\delta}^{(1)}$ is the coefficient of $\delta (1-x)$ in the 
two-loop quark-quark splitting function. 
Unlike the DY case (\ref{D2DY}) two new constants can occur at the NNLL 
level in DIS, hence the consistency of the resummed and the two-loop 
coefficient functions does not completely specify $g_3^{\rm DIS}$. 
%
For that either an extension of the calculations of refs.~\cite
{sglue1,sglue2} to the next order, or the $\ln^2 N$ term of the 
three-loop coefficient function (fixing $g_{32}$ which involves
the combination $B_2 + 2D_{1,2}$) is required. An approximate result 
has been derived for the latter~\cite{NVnew} from the constraints of 
refs.~\cite{moms1,moms2,av99}. Within errors that result is consistent 
with $D_2^{\rm DIS} = 0$, but, for instance, also with 
$\, B_2 \, = \, - P_{q,\delta}^{(1)} + \xi \, \beta_{0} C_F\, $ for 
$\,\xi \,\simeq\, 8 \,\ldots\, 13$.

The integrals in eqs.~(\ref{dint})--(\ref{Dint}) are not well-defined,
as they involve the running coupling at arbitrarily low scales. This
feature leads to the poles at $\,(2)\, a_s \beta_0 \ln N = 1\, $ in 
eqs.~(\ref{g1n})--(\ref{g3n}) and their DY counterparts, a problem which
is usually dealt with by the `minimal-prescription' contour for the 
Mellin inversion \cite{CMNT}. Another option, followed in ref.~\cite
{av99} for the DIS case, is to re-expand eq.~(\ref{cNres}) in powers of
$\alpha_s$ and to keep only those terms $\,\alpha_s^n \ln^{2n+1-k}N$, 
$k = 1,\ldots\, l$  (i.e., the first $l$ `towers' of logarithms) which 
are completely fixed by the known terms in eq.~(\ref{GNexp}) and in 
$g_0$ of eq.~(\ref{cNres}). In connection with a two-loop result for 
$g_0$ the NLL and NNLL resummations lead to $l=4$ and $l=5$, 
respectively. General expressions for the first four towers can be 
found in eq.~(14) of ref.~\cite{av99}$^2$, the extension to higher
terms is straightforward if lengthy. After Mellin inversion the NNLL
resummation thus leads to the following prediction for the five leading 
$x\rightarrow 1$ terms of the three-loop (quark-antiquark annihilation)
DY coefficient functions for $d\sigma/dQ^2$ at $\mu_f^2 = \mu_r^2 =
Q^2$ (the other terms are fixed by renormalization-group constraints): 
\bea
\label{cq3x}
  c_{q,\rm DY}^{(3)}(x) &\! =\! &
    512\, C_F^3 \left[ \frac{\ln^5 (1-x)}{1-x} \right]_+
    - \left( \frac{7040}{9}\, C_F^2 C_A^{} - \frac{1280}{9}\,
       C_F^2 N_f^{} \right) \left[ \frac{\ln^4 (1-x)}{1-x} \right]_+
    \nonumber \\[1mm] & & \mbox{}
    + \left\{ \bigg( - 2048 - 3072\, \zeta_2^{} \bigg) \, C_F^3
    + \bigg( \frac{17152}{9} - 512\, \zeta_2^{} \bigg) \, C_F^2 C_A^{}
    + \frac{7744}{27}\, C_F^{} C_A^2
    \right. \nonumber \\ & & \left. \mbox{} \quad\quad
    - \frac{2560}{9}\, C_F^2 N_f^{}
    - \frac{2816}{27}\, C_F^{} C_A^{} N_f^{}
    + \frac{256}{27}\, C_F^{} N_f^2 \right\}
    \left[ \frac{\ln^3 (1-x)}{1-x} \right]_+ \,
    \nonumber \\[1mm] & & \mbox{}
    + \left\{ 10240\,\zeta_3^{}\, C_F^3
    + \bigg( -\frac{4480}{9} + 1344\, \zeta_3^{} + \frac{11264}{3}
      \zeta_2^{} \bigg) C_F^2 C_A^{}
    \right. \nonumber \\ & & \left. \mbox{} \quad\quad
    + \bigg( -\frac{28480}{27} + \frac{704}{3}\, \zeta_2^{} \bigg)
       C_F^{} C_A^2
    + \bigg( \frac{544}{9} - \frac{2048}{3}\, \zeta_2^{} \bigg)
       C_F^2 N_f^{}
    \right. \\ & & \left. \mbox{} \quad\quad
    + \bigg( \frac{9248}{27} - \frac{128}{3}\, \zeta_2^{} \bigg)
       C_F^{} C_A^{} N_f^{}
    - \frac{640}{27}\, C_F^{} N_f^2 \right\}
    \left[ \frac{\ln^2 (1-x)}{1-x} \right]_+ 
    \nonumber \\[1.5mm] & & \mbox{}
    - \Big( 77949.50 -  5886.63\, N_f + 32.888\, N_f^2 + 4 A_3
      \Big) \left[ \frac{\ln (1-x)}{1-x} \right]_+ 
    + \ldots \:\: . 
    \nonumber 
\eea
Here $c_{q,\rm DY}^{(3)}(x)$ refers to $a_s = \alpha_s/(4\pi)$, as all 
other expansion parameters in this paper. Since $A_3$ is presently 
known just in the form (\ref{a3}), only the numerical result has been 
written down for the $[(1-x)^{-1}\ln (1-x)]_+$ term in eq.~(\ref{cq3x}).
In practice, however, the uncertainty of $\pm 12$ for $A_3$ is 
irrelevant here and in the corresponding higher-order terms. 

An improvement on the tower expansion for DIS in 
ref.~\cite{av99}\footnote
{Note that the third term of eq.~(15) has been misprinted in the 
 original preprint as well as the journal version of ref.~\cite{av99}:
 $16/27\, C_F^2 N_f$ has to be replaced by $16/27\, C_F N_f^2$, and
 $70/9\, C_F N_f^2$ by $280/9\, C_F^2 N_f$.}
will only be possible once $B_2^{\rm DIS}$ in eq.~(\ref{BD2DIS}) is 
exactly determined. However, the present results for $g_3^{\rm DIS}$
are directly relevant to the `physical' evolution kernels $K^N$ for the 
scaling violations of the non-singlet structure functions investigated 
beyond order $\alpha_s^2$ in refs.~\cite{NVnew,NV1}. Identifying all 
scales for brevity these kernels can be written as
\beq
\label{kns}
  \frac{dF_{a,\rm NS}^N}{d\ln Q^2} \: = \: \bigg( P^N(a_s) 
  + \frac{d \ln C_a^N(a_s)}{d a_s} \beta (a_s) \bigg) \: F^N_{a,\rm NS}
  (Q^2) \: \equiv \: K^N(a_s) \: F^N_{a,\rm NS}(Q^2) \:\: ,
\eeq
where $P^N(a_s)$ and $C_a^N(a_s)$ ($a = 1,\, 2,\, 3)$ are the moments 
of the non-singlet splitting functions and coefficient functions,
respectively. At large-$x\,$/$\,$large-$N$ eq.~(\ref{kns}) holds for 
the full structure functions up to corrections of order $1/N$. 
Inserting eqs.~(\ref{cNres}) and (\ref{GNexp}) into this equation
one arrives at the NNLL resummation of this kernel,
\bea
\label{kres}
  K_{a,\rm res}^N(a_s) & \! = \! & 
  \mbox{} - \ln N \, (A_1\,a_s + A_2\,a_s^2 + A_3\,a_s^3) \: - \: 
  \bigg( 1+ \frac{\beta_1}{\beta_0} a_s + \frac{\beta_2}{\beta_0} 
  a_s^2 \bigg)\, \lambda^2 \frac{dg_1^{}}{d\lambda} 
  \\ & & \mbox{} 
  - \Big( a_s \beta_0 + a_s^2 \beta_1 \Big)\, \lambda 
  \frac{dg_2^{}}{d\lambda} \: - \: a_s^2 \beta_0\, 
  \frac{d}{d\lambda} \Big( \lambda g_3^{}(\lambda)\Big) 
  \: + \: {\cal O}(a_s^3 (a_s \ln N)^n) \quad \nonumber
\eea
with $\lambda = a_s \beta_0 \ln N$, $A_i$ of eqs.~(\ref{a12}) and 
(\ref{a3}), and $g_i^{}$ from eqs.~(\ref{g1n})--(\ref{g3n}) with
$\mu_f = \mu_r = Q$. 
Note that the leading large-$N$ term of the quark-quark splitting 
function is given by $\ln N$ at all orders in $\alpha_s$ \cite{Ko89}.

Finally we briefly illustrate the numerical impact of the NNLL 
corrections. For these illustrations we choose $\mu_r^2 = \mu_f^2 = 
Q^2$, $N_f = 4$ and $\alpha_s = 0.2$. Depending on the precise value of 
$\alpha_s(M_Z^2)$, the latter number corresponds to scales between 
about 25 and 50 GeV$^2$, a range typical for fixed-target experiments 
both on DIS and the Drell-Yan process. The corresponding results for 
the functions $G^N_{\rm DIS}$ and $G^N_{\rm DY}$ are presented in 
fig.~1. Their convolutions with a schematic, but typical input are 
shown in fig.~2, where the minimal-prescription Mellin inversion 
\cite{CMNT} has been employed. The remaining ambiguity in the DIS case 
(discussed below eq.~(\ref{BD2DIS})) is illustrated by the results for 
$\, D_{2}^{\rm DIS} = 0\, $ and for $\, B_2 \, =\, - P_{q,\delta}^{(1)}
+ 13 \,\beta_{0} C_F\, $ (denoted by $\xi =13$ in the figures), the 
latter curves indicating the maximal NNLL effects consistent with the
available three-loop information \cite{moms1,moms2,av99,NVnew}. 
It is obvious from both figures that knowledge of the $\ln N (\alpha_s 
\ln N)^n$ and $(\alpha_s \ln N)^n$ terms \cite{sglue1,sglue2} alone is
not sufficient for reliably determining the functions $G_P^N$ and their 
impact after convolution even for rather moderate values of $N$ and $x$ 
(often denoted $\tau$ in the DY case). On the other hand the NNLL 
corrections presented in this paper are rather small over a wide range, 
e.g., less than 10\% and 20\% at $x \leq 0.85$ and $\sqrt x \leq 0.75$ 
for the DIS and DY results of fig.~2, respectively. 
This stabilization indicates that the soft-gluon exponents $G_P^N$ and 
their effects can now be reliably estimated for these processes. 
       
\vspace{1mm}
\noindent
{\bf Acknowledgements}\\[1mm]
It is a pleasure to thank E. Laenen and W.L. van Neerven, but 
in particular W. Vogelsang (who checked the non-$\zeta_2$ parts of 
eq.~(\ref{g3n}) using a conventional NLL technique) for useful 
discussions. 
The author has also profited from brief conversations with S. Catani 
and G. Sterman. 
This work has been supported by the European Community TMR research 
network `Quantum Chromodynamics and the Deep Structure of Elementary 
Particles' under contract No.~FMRX--CT98--0194.
\newpage
\newpage
\centerline{\epsfig{file=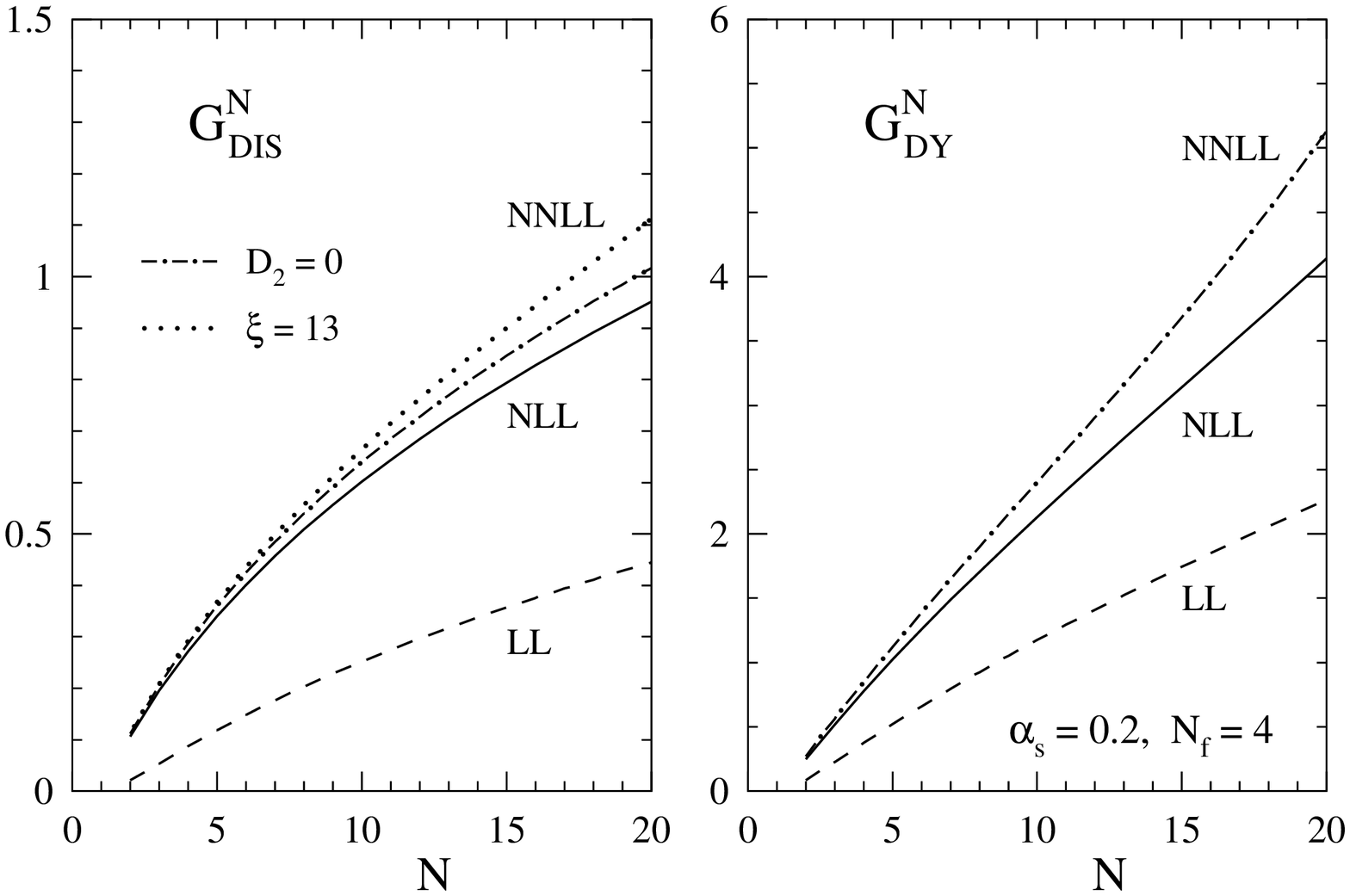,width=15cm,angle=0}}

\vspace{-3mm}
\noindent
{\bf Fig.~1:} The LL, NLL and NNLL approximations for the resummation
 exponents $G^N(Q^2)$ in eq.~(\ref{GNexp}) at $\mu_r^2 = \mu_f^2 = Q^2$ 
 for $\alpha_s(Q^2) = 0.2$ and four flavours. The two NNLL curves in 
 the DIS case indicate the present uncertainty as discussed below 
 eq.~(\ref{BD2DIS}).

\vspace{6mm}
\centerline{\epsfig{file=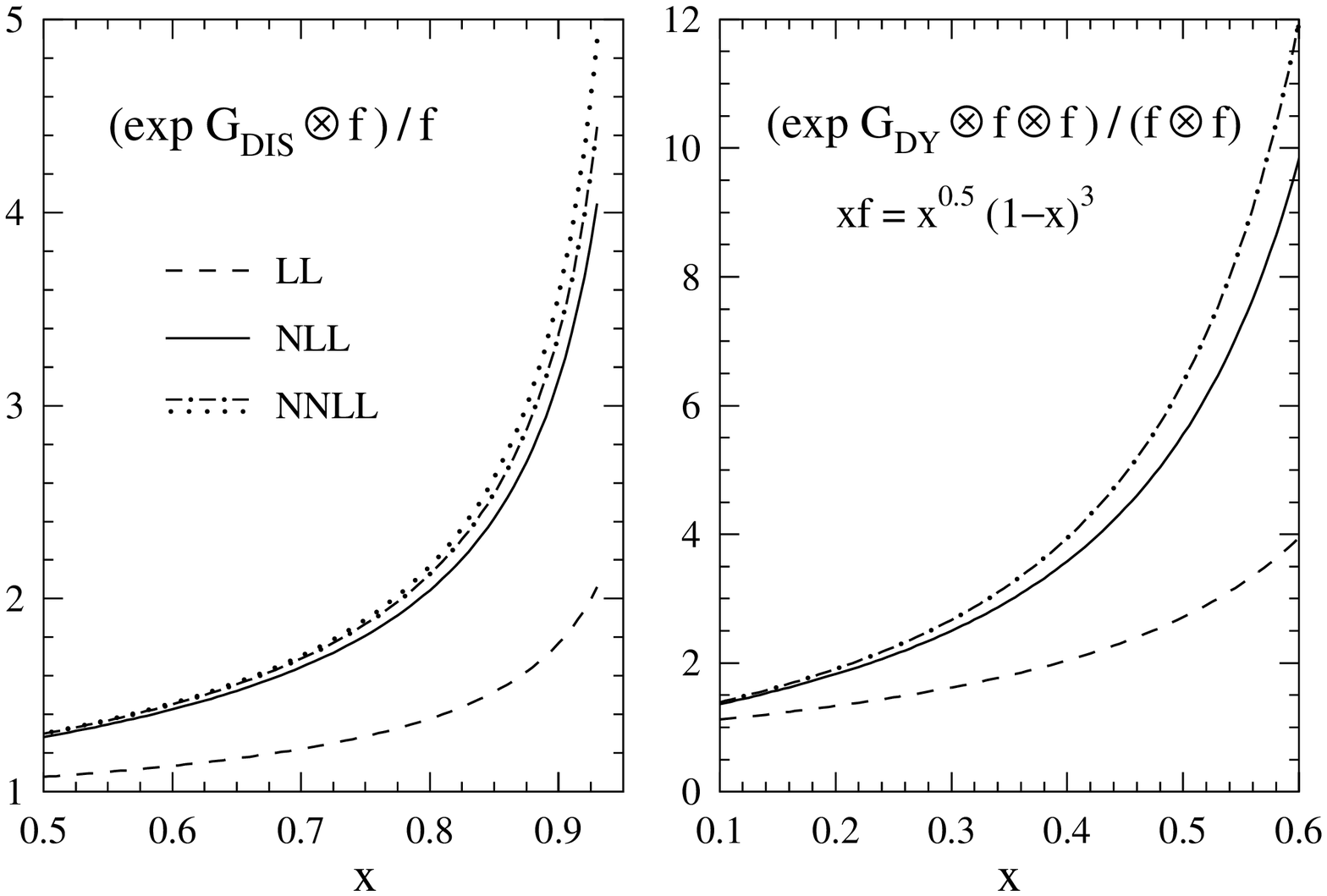,width=15cm,angle=0}}

\vspace{-3mm}
\noindent
{\bf Fig.~2:} The convolutions of the results shown in Fig.~1 with a 
 typical input shape. 

\end{document}